\documentclass[a4paper,12pt]{article}

\usepackage{amsmath,amssymb}
\usepackage[dvips]{graphics}
\usepackage{epsfig}
\usepackage{psfrag}
\newcommand{\be}{\begin{equation}}
\newcommand{\ee}{\end{equation}}
\def\bea{\begin{eqnarray}}
\def\eea{\end{eqnarray}}

\font\psyra=psyr at 9pt
\def\tauup{\mbox{\psyra t}}

\begin{document}

\begin{center}
{\Large \bf Extended Ho\v{r}ava Gravity with Physical Ground-State
Wavefunction} \vspace{20mm}

{\large Fu-Wen Shu\footnote{e-mail address: shufw@cqupt.edu.cn}}\\

\vspace{3mm} {\em  College of Mathematics and Physics, Chongqing
University of Posts and Telecommunications, Chongqing, 400065,
China}

\end{center}


\begin{abstract}
~We propose a new extended theory of Ho\v{r}ava gravity based on the
following three conditions: (i) UV completion, (ii) healthy IR
behavior and (iii) a stable vacuum state in quantized version of the
theory. Compared with other extended theories, we stress that any
realistic theory of gravity must have physical ground states when
quantization is performed. To fulfill the three conditions, we
softly break the detailed balance but keep its basic structure
unchanged. It turns out that the new model constructed in this way
can avoid the strong coupling problem and remains power-counting
renormalizable, moreover, it has a stable vacuum state by an
appropriate choice of parameters.
\end{abstract}

PACS: number(s): 04.60.-m, 05.10.Gg

\newpage
\section{Introduction}

Recently a new attempt to formulate a consistent and renormalizable
quantum theory of gravity has received extensive attention. This is
an ultraviolet(UV) renormalizable theory of gravity proposed by
Ho\v{r}ava in \cite{horava09}. Inspired by the perspectives existed
in the theory of dynamical critical systems and quantum criticality,
the proposal assumes that the space and time are anisotropic
\begin{equation}
x^i \rightarrow bx^i,\ \ \ t\rightarrow b^z t,
\end{equation}
where $z\geq 1$ is the dynamical critical exponent. In the UV regime
it has $z>1$. The theory will flow to $z=1$ in the infrared (IR)
region. The Lorentz invariance is obviously violated as $z>1$ but it
assumes that there is a foliated diffeomorphism invariance with
respect to the spatial sector\footnote{ Although the Lorentz
invariance has been verified experimentally at sufficiently large
scales, it is possible to have a Lorentz violation at high
energies\cite{jacobson,kostelecky,bailey,chadha}. This possibility
also has been partially confirmed in some experiments, see
\cite{chiow} and \cite{gamma} for examples.}. By adding higher order
spatial derivative terms into the Lagrangian it can reconcile the UV
divergence and make the theory renormalizable by power-counting. It
is this perspective that enables the proposal to attract a lot of
interests in recent literatures. These papers include from the
attempts at finding the classical solutions \cite{lu} to the
application to cosmology\cite{calcagni,kiritsis}, and other aspects
(see \cite{horava1} for an incomplete list). In principle, the
independent higher order terms which are allowed in the action seems
to be extremely large, leading to the theory lack of predictive
power. Ho\v{r}ava overcome this difficulty by introducing an
additional condition into the theory---the so called ``detailed
balance'', an idea borrowing from the condensed matter physics.

On the other hand, in Ho\v{r}ava's original proposal\cite{horava09},
it was argued that the Lorentz invariance can be recovered in the IR
limit where $z$ flows to $1$. In this limit the Einstein's theory
naturally appears assuming that a parameter $\lambda$ (a
dimensionless coupling measuring the breaking of the full
diffeomorphism group) also flows to $1$ in the same limit. However,
recent progress indicates that the theory exhibits a pathological
behavior at the low energies. Generally speaking, the pathologies
include the following two aspects: the strong coupling problem in
the IR fixed point\cite{cc} and the non-closure of constraint
algebra\cite{miao,farkas}. Essentially, these two pathologies have
the same origin. As pointed out in\cite{bps}, this is mainly due to
the fact that the breaking of general covariance by the preferred
foliation of space-time introduces a new scalar excitation. A recent
effort attempting to overcome these difficulties is an extended
theory of the non-projectable Ho\v{r}ava gravity proposed by Blas,
Pujol\`{a}s and Sibiryakov (BPS)\cite{bps2}. The key idea of this
extended theory (we denote it by BPS theory hereafter) is to improve
the IR behavior by breaking the ``detailed balance'' and introducing
a new $3$-vector and its higher derivatives into the Lagrangian. As
pointed out in \cite{ps}, this extension could still possess strong
coupling at low energies as we consider cubic or higher order
Lagrangian, but it is also possible to avoid the strong coupling if
higher derivative terms in the action become important below the
strong coupling energy scale\cite{bps2}.

So far it seems that the BPS model is an ideal theory of gravity
exhibiting healthy behavior at both high and low energies. However,
there are at least two obvious obstacles that prevent us from the
final theory. First, by giving up the ``detailed balance'', the
potential term in the action appears to include a large number of
terms and hence the number of the parameters needed in this model
would be very large, making the theory lack of predictive power.
Second, a well-defined quantized theory of the model constructed in
this way cannot be guaranteed in the sense that the model may have
unphysical ground states. Therefore, we should refine our model by
carefully selecting terms in the action so that the model has a
well-defined quantized theory. Meanwhile, to make the theory have
predictive power, the number of the parameters in the action should
be as less as possible. In this paper, we are paying our attention
to these problems and trying to construct our theory
of gravity based on the following three conditions:\\
 (i) UV
completion, in the sense that the candidate theory should be
renormalizable in the UV regime; \\
(ii) has a healthy IR behavior, namely, the theory should be free of ghost and does not have strong coupling;\\
(iii)can be well quantized in the sense that the theory has a stable
vacuum state (physical ground state).

In performing quantization of our model, we apply the stochastic
quantization method\cite{wu}, which is constructive through
stochastic differential equation, so that the question of whether a
stable vacuum (ground state) really exists or not can be easily
investigated and answered. Also it has the great advantage of no
need for gauge-fixing when applied to theories with gauge symmetry.
Its equivalence to path integral has been well proved in a lot of
literatures (see\cite{eg} for example).

The organization of the rest of the paper is as follows. In section
2, we start with a brief review of Ho\v{r}ava gravity and its
healthy extension. Section 3 focuses on the power-counting
renormalization analysis on our new model. Detailed study on the IR
behavior of the model is given in section 4, where we will show that
our model is free of the strong coupling problem. In section 5, we
pay our attention to the quantization of our theory using stochastic
quantization. We will show that the theory has a stable vacuum state
if $\lambda<1/3$. Conclusions and discussions are given in the last
section.

\section{Anisotropic theory of gravity}
For an anisotropic theory of gravity as suggested by Ho\v{r}ava, a
power-counting renormalizable action can be constructed by
considering the ADM decomposition of the space-time metric
\be ds^2= - N^2  dt^2 + g_{ij} (dx^i - N^i dt) (dx^j - N^j dt)\,,
\ee
where $N$ and $N_i$ are the lapse and shift functions respectively.
The spatial metric $g_{ij}$ with $i,j=1,2,3$ for $(3+1)-$dimensional
spacetimes has a Euclidean signature. For $z=3$ theory, a generic
action to be power-counting renormalizable is of the
form\cite{horava09,bps}
\begin{equation}\label{action}
S=\int dt d^3x
\sqrt{g}N(\frac2{\kappa^2}\mathcal{L}_K-\kappa^2\mathcal{L}_V),
\end{equation}
where $g$ denotes the determinant of the spatial metric $g_{ij}$.
The kinetic term is given by
\be \label{kinetic}\mathcal{L}_K \equiv \mathcal{O}_K=K_{ij} K^{ij}
- \lambda K^2=K_{ij}G^{ijkl}K_{kl} \,, \ee
where $K_{ij}$ is defined by
\be K_{ij} = \frac{1}{2N} (\dot g_{ij} - \nabla_i N_j - \nabla_j
N_i)\,, \ee
and $K\equiv K_i^i$. The symbol $G^{ijkl}$ is the generalized De
Witt metric
\be G^{ijkl}=\frac12(g^{ik}g^{jl}+g^{il}g^{jk})-\lambda
g^{ij}g^{kl}, \label{dewitt}\ee
with $\lambda$ a dimensionless free parameter.

The potential term in \eqref{action} which satisfies both the
power-counting renormalizable condition and
foliation-diffeomorphisms is of the form
\be \label{potential}\mathcal{L}_V
=\sum_{n=2}^6D_n(\Lambda,g_{ij},R_{ij},\nabla_i R_{jk},\cdots) \,,
\ee
where $D_n (n=2,\cdots,6)$ denote all possible scalars constructed
of $\Lambda,g_{ij},R_{ij},\nabla_i R_{jk},\cdots$ with the same
dimension $n$ and spatial parity. In particular, $D_2$ is of the
form $-(R-2\Lambda)$ to have a GR limit. A possible term of $D_3$ is
$\epsilon^{ijk}\nabla_iR_{jk}$, but it is excluded by spatial
parity. $D_4$ may include terms like $R_{ij}R^{ij}$, $\Delta R$
etc.. While the only possible term with spatial parity for $D_5$ is
$\epsilon^{ijk}R_{il}\nabla_jR_k^l$. The highest dimension allowed
by renormalizable condition is $6$ and all terms with dimension $6$
constitutes $D_6$ which has $R_{ij}R^{jk}R^i_{k}$,
$\nabla_iR_{jk}\nabla^{i}R^{jk}$ and $R\Delta R \cdots$ as its
ingredients.

Recent progress on Ho\v{r}ava gravity turns out, however, that the
action constructed as \eqref{action} does not have a healthy
infrared behavior---it suffers from a strong coupling problem due to
the violation of the diffeomorphisms for the full spacetimes. A
possible way out of this difficulty was recently suggested in
\cite{bps} by introducing the potential a set of terms which are
constructed from a $3$-vector
$$
\mathcal{E}_i\equiv\frac{\partial_i N}{N}.
$$
Explicitly, the extra terms of potential is
\bea \label{epotential}\delta\mathcal{L}_V =-\alpha
\mathcal{E}_i\mathcal{E}^i+\beta
(\mathcal{E}_i\mathcal{E}^i)^2+\gamma\mathcal{E}_i\Delta
\mathcal{E}^i+\delta\mathcal{E}_i\mathcal{E}_jR^{ij}+\cdots\,\eea
where $\alpha,\beta,\gamma, \delta$ are coupling constants and
ellipse represents all other possible terms constructed from
$\mathcal{E}_i$ and its covariant derivatives but the following
conditions should be satisfied\cite{bps}: (a) power-counting
renormalizability, this is equivalent to require that all the terms
should have dimensions no more than $6$, (b) spatial parity and, (c)
time-reversal invariance. Action constructed in this way turns
out\cite{bps} to be renormalizable by power-counting and free of
strong coupling problem.

So far it seems that we have a good theory of gravity by
constructing the gravity action in the way given above. However, as
mentioned in the last section, there are at least two obvious
obstacles that prevent us from the final result: (i) the potential
term in the action \eqref{action} appears to include a large number
of terms and hence the number of the parameters needed in this model
would be very large, making the theory lack of predictive power, and
(ii) a well-defined quantized theory of the model constructed in
this way cannot be guaranteed in the sense that the model may have
unphysical ground states (we will show this explicitly in section
\ref{sec5}). Motivated by these considerations, we refine our model
by carefully selecting terms in the action so as to the model has a
well-defined quantized theory. Meanwhile, to make the theory have
predictive power, it is better to has parameters as less as possible
in the action. Ref. \cite{shu} shows that for Ho\v{r}ava gravity it
is possible to have a physical ground state, and that the detailed
balance structure plays an important role in achieving so. For this
reason, we keep the basic structure of Ho\v{r}ava's theory, but add
terms that contribute to the IR behavior to softly break it.
Explicitly, the action is of the form
\be S=\int d^3xdt\sqrt{g}N(\frac2{\kappa^2}K_{ij}G^{ijkl}K_{kl}
-\frac{\kappa^2}8
E^{ij}G_{ijkl}E^{kl}+\alpha\mathcal{E}_i\mathcal{E}^i
),\label{exaction} \ee
where $E^{ij}$ is given by
\be \sqrt{g}E^{ij}=\frac{\delta W}{\delta g_{ij}}\,, \ee
with
\be \label{w}W=\mu_1\int \omega_3+\mu_2\int
d^3x\sqrt{g}(R-2\Lambda_W)\,, \ee
where
\be \omega_3=Tr(\Gamma\wedge
d\Gamma+\frac23\Gamma\wedge\Gamma\wedge\Gamma)\,, \ee
and $\mu_i (i=1,2)$ are coupling constant with scaling dimensions
$[\mu_i]_s=i-1$ and $[\Lambda_W]_s=2$. The model \eqref{exaction} is
largely simplified and only very limit parameters are needed. It is
also obviously renormalizable by power counting and is free of
strong coupling problem since the main contribution of $\delta
\mathcal{L}_V$ in \eqref{epotential} in the IR limit comes from
$\mathcal{E}_i\mathcal{E}^i$. Meanwhile, the theory
\eqref{exaction}, when a proper choice of parameters are made, can
be well quantized at least in the context of stochastic quantization
as will see below. We will give more details in the following
sections.

\section{UV completion}
In this section we would like to show, in an explicit way, that the
extended theory is power-counting renormalizable. To make the
analysis more convenient, one rewrites the action \eqref{action} in
a more explicit form
\be\label{3action}  S=\int
d^3xdt\sqrt{g}N(\frac2{\kappa^2}K_{ij}G^{ijkl}K_{kl}
-\sum_{a=2}^{6}\lambda_a \mathcal{O}_a )\,, \ee
where
\be \lambda_6 \equiv \frac{\kappa^2\mu_1^2}2,\ \ \lambda_5 \equiv
-\frac{\kappa^2\mu_1\mu_2}2,\ \ \lambda_4 \equiv
\frac{\kappa^2\mu_2^2}{8},\ \ \lambda_2 \equiv \frac{
\Lambda_W\lambda_4}{3\lambda-1}\label{constant} \ee
and
\bea
\nonumber\mathcal{O}_2&=&R-3\Lambda_W-\hat{\alpha}\mathcal{E}_i\mathcal{E}^i,\
\ \
\mathcal{O}_4=R_{ij}R^{ij}-\frac{1-4\lambda}{4(1-3\lambda)}R^2\\
\mathcal{O}_5&=&\epsilon^{ijk}R_{il}\nabla_jR_k^l,\ \ \
\mathcal{O}_6=C_{ij}C^{ij},
 \eea
where $\hat{\alpha}\equiv \frac{\alpha}{\lambda_2}$ and $C_{ij}$ is
the Cotton tensor, defined by
\be C^{ij}\equiv \epsilon^{ikl}\nabla_k\left(R^j{}_l
-\frac14R\delta_l^j\right).\label{cotton} \ee
The scaling dimensions of the coefficients of terms in the action
\eqref{3action} are
$$
[\kappa^2]_s=z-3,\ \ [\lambda_a]_s=z+3-a,\ \ [\hat{\alpha}]_s=0.
$$
In the context of Ho\v{r}ava-Lifshitz gravity, the dynamical
critical exponent in the UV regime is $z=3$, implying that
$\mathcal{O}_K$ and $\mathcal{O}_6$ are marginal terms and other
terms are relevant. Hence the theory is renormalizable by power
counting. While in IR regime, where the dynamical critical exponent
is flowed to $z=1$, we find only $\mathcal{O}_K$ and $\mathcal{O}_2$
are relevant with $\mathcal{O}_4$ marginal, in this limit we reach
the low-energy effective theory of gravity (up to the
$\mathcal{O}_4$ term).

\section{IR behavior}
\label{sec4}
To see the IR behavior of the Ho\v{r}ava theory, we investigate the
quadratic Lagrangian of \eqref{3action} by introducing the scalar
perturbations of the metric. By adopting the same gauge as the one
used in \cite{ps}, we obtain the scalar perturbations of metric
\bea N=e^{\phi(t,\vec{x})},\ \ \ \ \ N_i=\partial_i B(t,\vec{x}),\ \
\ \ g_{ij}=e^{2\psi(t,\vec{x})}\delta_{ij}. \label{pert metric} \eea
Substituting \eqref{pert metric} into the action \eqref{3action} and
integrating by part we obtain the following quadratic terms
\bea
\mathcal{O}_K^{(2)}&=&3(1-3\lambda)\dot{\psi}^2-2(1-3\lambda)\dot{\psi}\Delta
B+(1-\lambda)(\Delta B)^2\\
\mathcal{O}_2^{(2)}&=& -4\phi \Delta \psi+2(\partial \psi)^2-\frac32\Lambda_W(\phi+3\psi)^2+\hat{\alpha}\phi \Delta \phi\\
\mathcal{O}_4^{(2)}&=& \frac{2(\lambda-1)}{1-3\lambda}\psi \Delta^2
\psi,\ \ \ \mathcal{O}_5^{(2)}= \mathcal{O}_6^{(2)}=0.
 \eea
It is obvious that the above quadratic Lagrangian reduces to those
obtained in \cite{ps} once we set $\Lambda_W=0$. Following \cite{ps}
the momentum constraints can be obtained by varying the quadratic
action with respect to $B$,
\bea \label{mconstr}\Delta B=\frac{3\lambda-1}{\lambda-1}\dot{\psi}.
 \eea
Similarly, varying the quadratic action with respect to $\phi$ we
obtain
\be \label{hconstr1}4\Delta \psi+3\Lambda_W
(\phi+3\psi)-2\hat{\alpha}\Delta \phi=0. \ee
To solve the constraint \eqref{hconstr1} we assume that
$\hat{\alpha}=-2/3$, then it yields
\be \phi=-3\psi. \label{hconstr2} \ee
The action for the extra scalar mode of the theory can be obtained
by substituting the constraints \eqref{mconstr} and \eqref{hconstr2}
into the quadratic Lagrangian
\be \label{m action} S^{(2)}=-\int d^3xdt
\left[\frac{2}{\kappa^2}\frac1{c_{\psi}^2}\dot{\psi}^2-\frac{2(\lambda-1-2\Lambda_W)\lambda_4}{3\lambda-1}(\partial
\psi)^2\right], \ee
where $c_{\psi}^2=\frac{1-\lambda}{3\lambda-1}$ is the speed of
sound for the mode $\psi$. It is straightforward from \eqref{m
action} that the dispersion relation of the propagating mode is
\be \label{dispersion} \omega^2=-\Big(\kappa^2 c_{\psi}^2
\frac{2(\lambda-1-2\Lambda_W)\lambda_4}{3\lambda-1}\Big)k^2\ee
From the quadratic action \eqref{m action} we see that the ghost can
be avoided by requiring $c_{\psi}^2<0$. This imposes a constraint on
$\lambda$
\be \label{constraint1} \frac{3\lambda-1}{\lambda-1}>0, \ee
implying $\lambda>1$ or $\lambda<1/3$. On the other hand, from the
dispersion relation \eqref{dispersion} the only way to avoid
exponential instabilities of the propagating mode $\psi$ is
\be \label{constraint2} \frac{\lambda-1-2\Lambda_W}{3\lambda-1}>0,
\ee
assuming $\lambda_4=\kappa^2\mu_2^2/8>0$. As \eqref{constraint1} is
satisfied this can be easily fulfilled by requiring
\be\label{condition} \frac{\Lambda_W}{3\lambda-1}<0, \ee
which is equivalent to require $\Lambda_W<0$ for $\lambda>1$ or
$\Lambda_W>0$ for $\lambda<1/3$.

The above analysis shows that, at least for quadratic action the
theory is free of strong coupling problem and exhibits a healthy IR
behavior as some conditions are fulfilled.

\section{Quantization of the theory}\label{sec5}
Recently most works on Ho\v{r}ava's gravity focus on the IR behavior
of the theory and try to refine the model by removing the
pathological behavior of the extra mode, as mentioned in the last
section. However, there is another most fundamental question should
be paid more attention, namely, whether the theory can really be
quantized in a consistent and non-perturbative manner? If yes,
whether this will put any constraint(s) on the parameters appearing
in the action or not? In this section we will, following the work
\cite{shu}, make a detailed analysis of these questions by using the
stochastic quantization.

\subsection{Brief review of stochastic quantization}
In this subsection, we give a brief survey of the stochastic
quantization. Generally speaking, the stochastic quantization can be
performed in the following steps:  (1) Transforming the action to
Euclidean version via an analytic continuation to imaginary time;
(2) Introducing a fictitious time to the system through which the
evolution of fields under random walk can be described. The
evolution equation is known as the Langevin equation; (3) Defining
the $n$-point correlation functions by taking averages over the
random noise field with a Gaussian distribution; (4) Identifying the
equal time correlators for the field with the corresponding quantum
Green's functions as the fictitious time approaches infinity. For
stochastic quantization, the key point is that the system is assumed
to be equilibrium for large fictitious time. In other words, the
Euclidean action is assumed to be bounded from below. The most
convenient way to see this point is to investigate the Fokker-Planck
equation \cite{fp} \cite{ili} associated with the equations
describing the stochastic dynamic of the system.

As an example, let us consider a free scalar field $\phi(x)$. As
mentioned we introduce a fictitious time $\tau$. Then the Langevin
equation, which describes the evolution of the system under random
motion, is given by
\begin{equation}
  \label{langevin_sc}
  \frac{\partial\,\phi(x,\tau)}{\partial \tau} = -\frac{\delta  S_{E}}{\delta \phi}+\eta(x,\tau),
\end{equation}
with $S_{E}$ the Euclidean action
\begin{equation}
S_{E}[\phi]=\int d^{d}x\, \left(\frac{1}{2}
(\partial\phi)^{2}+\frac{1}{2} m_{0}^{2}\,\phi^{2}(x)\right).
\label{scalar action}
\end{equation}
The white Gaussian noise $\eta$ in \eqref{langevin_sc} satisfies
\be \label{white n} <\eta (x,\tau)> = 0 \ \ \, <\eta (x_1,\tau_1)
    \eta(x_2,\tau_2)>=  2\, \delta ( \tau_1 - \tau_2)
  \delta^d ( x_1 - x_2 ),
\ee
The $n$-point correlation function is define as
\be
<\phi_{\eta}(x_1,\tau_1)\ldots\phi_\eta(x_k,\tau_k)>=\frac{\int\,\mathcal{D}[\eta]
\phi_{\eta}(x_1,\tau_1)\ldots\phi_\eta(x_k,\tau_k)\exp\biggl[-\frac{1}{4}
\int d^{d}x \int d\tau\,\eta^{2}(\tau,x)\bigg]}
{\int\,\mathcal{D}[\eta]\exp\biggl[-\frac{1}{4} \int d^{d}x \int
d\tau\,\eta^{2}(\tau,x)\bigg]}.
 \ee
Identifying this correlation function with the corresponding quantum
Green's functions as the fictitious time approaches infinity, i.e.,
\be
\lim_{\tau\to\infty}<\phi_{\eta}(x_1,\tau_1)\ldots\phi_\eta(x_k,\tau_k)>\mid_{\tau_1=\cdots=\tau_k=\tau}=
<\phi_{\eta}(x_1)\ldots\phi_\eta(x_k)>, \ee
In particular, for the action given by \eqref{scalar action}, it is
easy to show that the equal time two-point correlation function in
phase space is given by
\be
<\phi(\tau,k)\phi(\tau,k')>=(2\pi)^d\delta^{d}(k+k')\frac{1}{(k^{2}+m_{0}^{2})}\biggl(1-\exp\left(-2\tau
(k^{2}+m_{0}^{2})\right)\biggr). \ee
Therefore, the Euclidean two-point function is recovered as
$\tau\rightarrow\infty$.

On the other hand, the existence of an equilibrium state can be
proved or disproved by studying the corresponding Fokker-Planck
equation associated with the Langevin equation. This is given by
\begin{equation}\label{Fokker_Planck}
  \frac{\partial P(\phi,\tau)}{\partial\tau} = \frac{\partial}{\partial \phi}
  \left(\frac{\partial}{\partial \phi}+\frac{\partial S_E}{\partial\phi}\right)
  P(\phi,\tau),
\end{equation}
where $P$ is the probability density which satisfies the
normalization condition
\begin{equation}
  \int d\phi\, P(\phi,\tau) = 1.
\end{equation}
 Solving the Fokker-Planck equation \eqref{Fokker_Planck} for given
$S_E$ one can obtain the probability density. An equilibrium state
of a system is supposed to have a positive and finite $P$.
\subsection{Quantization of BPS model}
Although the extended Ho\v{r}ava gravity \cite{bps} succeeds in
avoiding the strong problem of Ho\v{r}ava's original scheme, it is
not guaranteed that the theory can be quantized in a consistent way
and that it has a well-defined physical ground state. In this
subsection, we would point out that the BPS model in its original
form may have unphysical ground states since the candidate
ground-state function is not always normalizable.

We start with the BPS action
\be \label{bps action} S_{BPS}=\int d^3xdt
\left[\frac2{\kappa^2}\mathcal{L}_K-\kappa^2(\mathcal{L}_V+\delta
\mathcal{L}_V)\right] \ee
where $\mathcal{L}_K$, $\mathcal{L}_V$ and $\delta\mathcal{L}_V$ are
given, respectively, by \eqref{kinetic}, \eqref{potential} and
\eqref{epotential}. Performing a wick rotation $t\rightarrow
i\tauup$ we obtain the Euclidean action of \eqref{bps action}, which
is denoted by $S^{bps}_E$ hereafter. Then the Langevin equation of
the BPS theory is \cite{shu}
\bea
\begin{cases}
\dot{N}=-\frac{1}{\sqrt{g}}\frac{\delta S^{bps}_E}{\delta N}+\eta,\\
\dot{N_i}=-\frac{1}{\sqrt{g}}\frac{\delta S^{bps}_E}{\delta N^i}+\zeta_ae^a{}_i,\\
\dot{\mathcal{E}_i}=-\frac{1}{\sqrt{g}}\frac{\delta S^{bps}_E}{\delta \mathcal{E}^i}+\sigma_ae^a{}_i,\\
\dot{g}^{I}=-\mathcal{G}^{IJ}\partial_{J}S^{bps}_{E}+\xi^AE_A{}^I,
\end{cases}\label{langevin} \eea
where the dot represents derivative with respect to the fictitious
time $\tau$ and following notations have been introduced:
$$
g_{ij}\equiv g^I,\ \ \ \ \mathcal{G}^{IJ}\equiv \mathcal{G}_{ijkl},
\ \ \ \ \partial_I S^{bps}_{E}\equiv \frac1{\sqrt{g}}\frac{\delta
S^{bps}_{E}}{\delta g_{ij}}.
$$
In Eq. \eqref{langevin}, we also have introduced vielbein
\bea e_a{}^ie_b{}^jg_{ij}=\delta_{ab},\ \ \
E_A{}^IE_B{}^J\mathcal{G}_{IJ}=\delta_{AB},\\
e_a{}^ie_b{}^j\delta^{ab}=g^{ij},\ \ \
E_A{}^IE_B{}^J\delta^{AB}=\mathcal{G}^{IJ}. \eea
so that noises $\eta$, $\zeta_a$, $\sigma_a$ and $\xi^A$ are
Gaussian and the following relations hold \cite{shu}
\bea
&&<\eta(x,\tau)>=0,\ \ <\zeta^a(x,\tau)>=0,\ \ <\sigma^a(x,\tau)>=0,\ \ <\xi^A(x,\tau)>=0,\\
&&<\eta(x,\tau)\eta(y,\tau')>=2\delta(x-y)\delta(\tau-\tau'), \\
&&<\zeta^a(x,\tau)\zeta^b(y,\tau')>=2\delta^{ab}\delta(x-y)\delta(\tau-\tau'), \\
&&<\sigma^a(x,\tau)\sigma^b(y,\tau')>=2\delta^{ab}\delta(x-y)\delta(\tau-\tau'), \\
&&<\xi^A(x,\tau)\xi^B(y,\tau')>=2\delta^{AB}\delta(x-y)\delta(\tau-\tau').
\eea
(Here $x$ stands for Euclidean coordinates $(x^i,\tauup)$.) The
correlation functional then can be defined with respect to $\eta$,
$\zeta^a$, $\sigma_a$ and $\xi^A$ by
\bea \nonumber<\mathcal{F}(N,N_i,\mathcal{E}_i,g_I)>& \sim & \int
\mathcal{D}[\eta]\mathcal{D}[\zeta]\mathcal{D}[\sigma]\mathcal{D}[\xi]
\mathcal{F}(N,N_i,\mathcal{E}_i,g_I)\\&& \cdot
\exp\left[-\frac14\int d\tauup
d^3xd\tau\sqrt{g}N(\eta^2+\zeta^a\zeta_a+\sigma^a\sigma_a+\xi^A\xi_A)\right],
\label{correlation1}
\eea 
which is obviously Gaussian as desired.

As mentioned in the last subsection, a convenient way to study
whether the Langevin process \eqref{langevin} really converges to a
stationary equilibrium distribution is to explore the associated
Fokker-Planck equation,
\bea \frac{\partial Q(N,N^i,\mathcal{E}_i,g_I,\tau)}{\partial
\tau}=-\mathcal{H}_{FP}Q(N,N^i,\mathcal{E}_i,g_I,\tau). \label{fp2}
\eea 
Here we have introduced a new function $Q$ which is associated the
probability density through
\be Q(N,N^i,\mathcal{E}_i,g_I,\tau) \equiv
P(N,N^i,\mathcal{E}_i,g_I,\tau)e^{S_E/2},\label{definition} \ee
where the probability density functional is given by
\be P(N,N^i,\mathcal{E}_i,g_I,\tau)=\frac{\exp\left[-\frac14\int
d\tauup
d^3xd\tau\sqrt{g}N(\eta^2+\zeta^a\zeta_a+\sigma^a\sigma_a+\xi^A\xi_A)\right]}
{\int\mathcal{D}[\eta]\mathcal{D}[\zeta]\mathcal{D}[\sigma]\mathcal{D}[\xi]\exp\left[-\frac14\int
d\tauup
d^3xd\tau\sqrt{g}N(\eta^2+\zeta^a\zeta_a+\sigma^a\sigma_a+\xi^A\xi_A)\right]}.\label{prob}
\ee
The Fokker-Planck Hamiltonian $\mathcal{H}_{FP}$ in \eqref{fp2} is
of the form
\bea
\mathcal{H}_{FP}=a^{\dagger}a+g^{ij}a_i{}^{\dagger}a_j+g^{ij}\tilde{a}_i{}^{\dagger}\tilde{a}_j
+\mathcal{G}^{IJ}\mathcal{A}_I{}^{\dagger}\mathcal{A}_J.
\label{hamiltonian} \eea Here
$$
a=i\pi+\frac12\frac1{\sqrt{g}}\frac{\delta S^{bps}_E}{\delta N},\ \
a^i=i\pi^i+\frac12\frac1{\sqrt{g}}\frac{\delta S^{bps}_E}{\delta
N_i},\ \
\tilde{a}^i=i\tilde{\pi}^i+\frac12\frac1{\sqrt{g}}\frac{\delta
S^{bps}_E}{\delta \mathcal{E}_i},\ \ \mathcal{A}^I=i
\pi^I+\frac12\partial^IS^{bps}_{E},
$$
with $\pi$, $\pi^i$, $\tilde{\pi}^i$ and $\pi^I$, respectively, the
conjugate momenta of $N$, $N^i$, $\mathcal{E}_i$ and $g^I$:
$\pi=-i\frac1{\sqrt{g}}\frac{\delta}{\delta N}$,
$\pi^i=-i\frac1{\sqrt{g}}\frac{\delta}{\delta N_i}$,
$\tilde{\pi}^i=-i\frac1{\sqrt{g}}\frac{\delta}{\delta
\mathcal{E}_i}$, $\pi_I=-i
\partial_I$.
The time independent eigenvalue equation associated with Eq.
\eqref{fp2} is
\bea \mathcal{H}_{FP} Q_n(N,N^i,\mathcal{E}_i,g_I,\tau)=E_n
Q_n(N,N^i,\mathcal{E}_i,g_I,\tau). \eea
The solutions of Eq. \eqref{fp2} lead to the probability density
\bea
P(N,N^i,\mathcal{E}_i,g_I,\tau)=\sum_{n=0}^{\infty}a_nQ_n(N,N^i,\mathcal{E}_i,g_I)
e^{-S^{bps}_{E}/2-E_n\tau}.\label{prob2}
\eea
From \eqref{prob2} we show that the theory will approach an
equilibrium state $Q_0(N,N^i,\mathcal{E}_i,g_I)=e^{-S^{bps}_{E}/2}$
for large $\tau$ if and only if all $E_n>0$ ($n>0$ and with
$E_0=0$). This is equivalent to find the condition(s) under which
the Fokker-Planck Hamiltonian \eqref{hamiltonian} is non-negative
definite. Following the analysis made in \cite{shu} we show that
this can be fulfilled by requiring a positive definite De Witt
metric $\mathcal{G}^{IJ}$, i.e., $\lambda< 1/3$. The theory then
approaches an equilibrium
\bea P_0(N,N^i,\mathcal{E}_i,g_I)\equiv\lim_{\tau\rightarrow \infty}
P(N,N^i,\mathcal{E}_i,g_I,\tau)=a_0 e^{-S^{bps}_{E}}, \label{prob1}
\eea 
where
\be \label{normalization const}a_0=\frac1{\int
\mathcal{D}[N]\mathcal{D}[N_i]\mathcal{D}[\mathcal{E}_i]\mathcal{D}[g_I]e^{-S^{bps}_E(N,N^i,\mathcal{E}_i,g_I)}},
\ee
is the normalization constant. Note that the stationary candidate of
equilibrium state $P_0$ in \eqref{prob1} is far from a genuine
physical ground state. In other words, the normalization constant
$a_0$ in \eqref{normalization const} is not guaranteed to be finite.
It follows from \eqref{normalization const} that the normalizable
ground state is achieved by requiring a positive definite Euclidean
action $S^{bps}_E$. While from \eqref{bps action} we see the action
$S^{bps}_E$ is not always positive definite, implying that some
unphysical ground states appear. To cure this problem more
constraints have to be imposed on the potential terms.

\subsection{Stochastic quantization of our model}

In this subsection we would like to propose a possible prescription
for removing the unphysical vacuum state. Inspired by the result of
\cite{shu}, we found a possible way out is to keep the basic
structure of ``detailed balance''. However, there are a lot of
literatures(see \cite{cc} for example) show that the strict
``detailed balance'' will lead to a catastrophe of the theory---the
strong coupling problem as mentioned in the previous part of the
paper. To avoid the strong coupling, we have to violate the detailed
balance structure. To coordinate these two apparently incompatible
conditions smoothly, on one hand, we softly break the detailed
balance, on the other hand, we keep the basic structure of the
detailed balance. This leads to our extended action \eqref{exaction}
of Ho\v{r}ava gravity. This action violates the detailed balance by
introducing an extra term $\mathcal{E}_i\mathcal{E}^i$ whose
presence cures the strong coupling problem as analysed in
Sec.\ref{sec4}. Meantime, it keeps the basic structure of detailed
balance which leads to a cure of the unphysical ground states as
will see below.

To see this explicitly, we write down the Euclidean action of our
model \eqref{exaction},
\be S_E=\int d^3xd\tau\sqrt{g}N(\frac2{\kappa^2}K_{ij}G^{ijkl}K_{kl}
-\frac{\kappa^2}8
E^{ij}G_{ijkl}E^{kl}+\alpha\mathcal{E}_i\mathcal{E}^i
),\label{eucaction} \ee
Repeating the procedures given in the last subsection we can
quantize the theory using the stochastic quantization, and, similar
to the case of BPS theory, we obtain the following solution of the
Fokker-Planck equation
\bea
P(N,N^i,\mathcal{E}_i,g_I,\tau)=\sum_{n=0}^{\infty}a_nQ_n(N,N^i,\mathcal{E}_i,g_I)
e^{-S_{E}/2-E_n\tau},\eea
where $S_E$ is given by \eqref{eucaction}. Therefore, the theory
will approach an equilibrium state
$Q_0(N,N^i,\mathcal{E}_i,g_I)=e^{-S_{E}/2}$ for large $\tau$ as long
as the De Witt metric $\mathcal{G}^{IJ}$ is positive definite, or
equivalently, $\lambda< 1/3$. The candidate equilibrium state of the
theory is
\bea P_0(N,N^i,\mathcal{E}_i,g_I)\equiv\lim_{\tau\rightarrow \infty}
P(N,N^i,\mathcal{E}_i,g_I,\tau)=a_0 e^{-S_{E}},\label{gs}
\eea 
where again
\be \label{normalization const1}a_0=\frac1{\int
\mathcal{D}[N]\mathcal{D}[N_i]\mathcal{D}[\mathcal{E}_i]\mathcal{D}[g_I]e^{-S_E(N,N^i,\mathcal{E}_i,g_I)}},
\ee
is the normalization constant. As mentioned in the last subsection,
the key to obtain a stable vacuum state (or physical ground state)
is that the Euclidean action in \eqref{normalization const1} must be
positive definite. In our model this can be achieved by requiring
that both the De Witt metric and $\alpha$ are positive definite.
Explicitly, we rewrite the action \eqref{eucaction} as
\be S_E=\int
d^3xd\tau\sqrt{g}N\left[\frac2{\kappa^2}\mathcal{G}^{IJ}(K_IK_J-\frac{\kappa^4}{16}
E_IE_J)+\alpha g^{ij}\mathcal{E}_i\mathcal{E}_j
\right],\label{eucaction1} \ee
where $E_I=\partial_I W$ with $W$ given by \eqref{w}. Therefore,
$S_E$ is positive definite for $\lambda<1/3$ and $\alpha>0$. In Sec.
\ref{sec4} we have chosen $\alpha=-\frac23\lambda_2$ with
$\lambda_2$ is defined in \eqref{constant}. It is straightforward to
show that the condition to have $\alpha>0$ is
$$
\frac{\Lambda_W}{3\lambda-1}<0.
$$
This is precisely the condition \eqref{condition} with which the
theory is free of the strong coupling problem. This condition is
equivalent to require $\lambda<1/3$ for $\Lambda_W>0$. As a
consequence, the state \eqref{gs} is indeed a physical ground state
if $\lambda<1/3$.

\section{Conclusions and discussions}
Based on three conditions:  (i) UV completion, (ii) healthy IR
behavior and (iii) a stable vacuum state, we have constructed a new
extension of the Ho\v{r}ava's gravity. In some sense, this model is
an improvement of the BPS model by imposing an extra
constraint---the condition with which the theory has a stable vacuum
---on the theory. This is achieved by keeping the basic ``detailed
balance'' structure but adding the terms curing the IR pathologies
in the action. There are at least three merits when construct
theories in this way: First, it puts strong constraints on the
number of the allowed terms in the action, hence makes the theory
has predictive power; Second, it makes the Euclidean action of the
theory bounded from below when $\lambda<1/3$ is fulfilled. This is a
key condition to have a stable vacuum state for theories when we are
performing stochastic quantization or path integral quantization.
Third, it provides a possible way in avoiding the strong coupling
problem at low energies. Indeed, our analyses made in this paper
show that the theory constructed in this way can fulfill all the
three conditions mentioned above assuming the parameter $\lambda$
satisfies some conditions in different energy scales.

One point deserves further investigation is to check whether our
model can really avoid the strong coupling problem when we are
expanding the Lagrangian to higher order. Although the present paper
show that the theory exhibits a healthy IR behavior for the
quadratic Lagrangian, this is not guaranteed for higher order
Lagrangian. This is equivalent to check if there is a new scale
other than the Planck scale for suppressing the higher derivative
terms so that these terms become important before the strong
coupling appears \cite{bps2}. Meanwhile, it is worthy of further
study on the Hamiltonian formalism of our model so as to find the
constraint structure of the theory.

\section*{Acknowledgment}

The author would like to thank the hospitality of the University of
Utah where a part of this work was completed. This work was
partially supported by the NNSF key project of China under grant No.
10935013, the Natural Science Foundation Projects of CQ CSTC under
grant No. 2009BB4084 and the Scientific Research Foundation for the
Returned Overseas Chinese Scholars, State Education Ministry.

\end{document}